# Self-supported bulk MXene electrodes for electrochemical hydrogen applications

Rebeca Miyar[1], Bar Favelukis[2], Eva B. Mayer[3], Manoj Prabhakar[1], Yug Joshi[1], Gerhard Dehm[1], Jochen M. Schneider[3], Maria Jazmin Duarte[1], Barak Ratzker[1], Maxim Sokol[2,*]

[1] Max Planck Institute for Sustainable Materials, Max-Planck-Straße 1, Düsseldorf, Germany
[2] Department of Materials Science and Engineering, Tel Aviv University, P.O.B 39040, Ramat Aviv 6997801, Israel
[3] Materials Chemistry, RWTH Aachen University, Kopernikusstr. 10, Aachen, Germany

* Corresponding author: sokolmax@tauex.tau.ac.il

## Abstract

MXenes are promising candidates for electrochemical applications due to their high conductivity, tunable surface chemistry, and catalytic potential. However, their use in bulk electrode form remains unexplored despite advantages such as higher current density and improved mechanical integrity. Herein, we present a methodology for the fabrication of self-supported vdW solid $Ti_3C_2T_z$ MXene electrodes, produced by cold compaction followed by vacuum heat treatment at 600 °C, which effectively removes interlayer confined water and stabilizes the bulk 3D structure. The resulting binder-free electrodes exhibit enhanced mechanical robustness along with structural and chemical stability in various electrolytes. The MXene electrodes demonstrate adequate HER activity while maintaining electrochemical stability over time, with minimal oxidation or changes in termination surface chemistry. This approach is scalable and cost-effective, overcoming limitations of nanoscale MXene architectures in electrochemical environments and offering a practical pathway toward MXene-based materials for sustainable hydrogen energy technologies.

## Body

Hydrogen is known as a clean and efficient energy carrier for future sustainable energy systems, driving research efforts into the development of more efficient and scalable applications for hydrogen production and storage[1,2]. In particular, electrochemical processes are an attractive approach since they can operate at room temperature and ambient pressure, with the potential to be directly coupled with renewable power sources[3]. However, their successful implementation in modern industry critically depends on the development of robust, cyclable, and highly conductive electrode materials[3,4].

MXenes are family of two-dimensional (2D) transition metal carbides, nitrides and carbonitrides[5,6]. They have emerged as promising candidates for electrochemical applications due to their high conductivity, hydrophilicity, and tunable surface chemistry, yet their practical integration into electrochemical systems remains limited[7–10]. MXenes have been mainly researched in the form of 2D nanoscale materials[5,11], or implemented in various 3D architectures[12–14] such as deposited thin films or coatings[15,16], flexible free-standing films[17] or composites bound with polymers[10,18]. These architectures suffer from restacking[6,18], low density[18], and poor mechanical stability, hindering their performance and long-term stability in hydrogen-related systems[11,18–20]. Here, bulk electrodes refer to self-supported, millimeter scale MXene bodies that contain no polymer binder or conductive additives. To date, MXene in the form of bulk electrodes has not been explored despite its potential scalability and advantages such as higher current density and improved mechanical integrity.

MXenes, like other 2D materials, exhibit strong van der Waals (vdW) forces between the layers[11]. These forces enable the consolidation of 2D materials into 3D vdW bulks by employing cold compaction under pressures of hundreds MPa. Studies have shown that various 2D materials could yield 3D bulk pellets with high packing density, by compaction at or near room temperature [21–23]. MXenes naturally retain some water between the layers after synthesis, which assist in enabling them to be readily compacted into robust 3D bulk form[22]. While compaction produces dense MXene pellets with adequate mechanical properties, they crucially lack stability in liquid electrolytes, disintegrating under electrochemical conditions. This lack of progress is largely linked to MXene instability in aqueous environments, where residual interlayer water facilitates ion penetration and structural degradation.

Herein, we introduce facile post-compaction vacuum heat treatment (HT) to produce bulk, binder-free, self-supported, electrochemically stable MXene electrodes. This method consists of cold-compaction of 2D MXene flakes into mechanically stable vdW solids, followed by a vacuum HT removing residual intercalated water while preserving structural integrity and surface chemistry. The removal of residual water enhances the vdW bonding and chemically stabilizes 3D MXene bulk electrodes in various solutions, showcasing promising electrochemical durability and mechanical integrity in environments relevant for electrochemical hydrogen storage and catalysis. The demonstrated proof of concept opens a new path toward implementation of MXene electrodes in sustainable energy systems.

A schematic illustration of the $Ti_3C_2T_z$ electrodes preparation procedure is presented in **Fig. 1a**, showcasing all process steps from synthesis to compaction and HT. Multilayered $Ti_3C_2T_z$ MXene was synthesized by removing the "A" layer from $Ti_3AlC_2$ MAX phase through chemical etching using hydrofluoric and hydrochloric acid[24], followed by washing, filtering and oven drying. The resulting multilayer MXene powder is shown in **Fig. 1b**. After compaction at room temperature under ~600 MPa the MXene formed dense pellets with a density of 2.6 g/cm$^3$. The cross section of a representative pellet fracture surface is presented in **Fig. 1c**. Note that some residual water in the synthesized MXene powder can assist in more effective compaction, as previously shown by Zhu et al.[22]. This compaction approach can be used for the shaping of pellets in different shapes, sizes and thicknesses (**Fig. S1.**).

The XRD characterization of the MXene along the different steps is shown in **Fig. 1d**. The patterns depict typical multilayer MXene with some minor peaks of residual $AlF_3$ originating from the MXene synthesis[25,26]. The uniaxial pressure applied during compaction orients the flakes, causing them to slide and bond by vdW forces due to their high surface area. This can be seen by the near disappearance of the {110} reflections in the XRD pattern while the basal reflections become sharper (**Fig. 1d**). As previously shown by Ghidiu et al.[27], this behavior is consistent with strong flake alignment under uniaxial pressure, where the flakes are oriented with their c-axis perpendicular the applied pressure. The resulting pellets were heat treated at 600 °C for 2 h under vacuum to remove residual water and stabilize the 3D bulk MXene. The HT temperature was selected since it is high enough to eliminate interlayer bound water but low enough to avoid any surface termination exchange or oxidation[28] and preserve the MXene structure[29]. Noticeable peak shifts due to water removal can be observed in the XRD patterns (**Fig. 1d**). After HT the pellet loses roughly 5% of its weight, resulting in a final density of 2.6 g/cm$^3$ when completely dry. The MXene electrode is mesoporous (roughly 20% porosity) with a specific surface area of ~37 m$^2$/g and average pore size of 3.8 nm as determined by Branauer-Emmett-Teller (BET) measurements (**Fig. S2.**). The porosity enables high surface area and possible applicability in gas phase reactions.

Preliminary testing of the 3D bulk MXene stability in solution was performed by immersing pellets at different stages in various water-based electrolytes (see insets in **Fig. 1d**). Cold-compacted samples would readily disintegrate in all solutions, while after vacuum HT at 600 °C they remain intact indefinitely (**Fig. S3a-b.**). After two weeks in KOH and $H_2SO_4$ there is some change in the color of the solution (yellowish hue) indicating that some oxidation has occurred (**Fig. S3c.**). The vacuum during HT ensures removal of any residual water, as pellets heat treated in Ar atmosphere remain stable only in neutral or acidic environments but disintegrate in alkaline solutions (**Fig. S3d.**), due to ion intercalation[30]. This indicates that vacuum is essential for removing confined water before oxidation or unintended termination changes can occur. Therefore, suggesting that residual interlayer water in the compacted MXene promotes ion and solvent penetration in such aqueous environments, causing swelling, delamination, and oxidative degradation. Only pellets after HT at 600 °C under vacuum retain their structural integrity over extended durations when immersed in various solutions (**Fig. S3e.**).

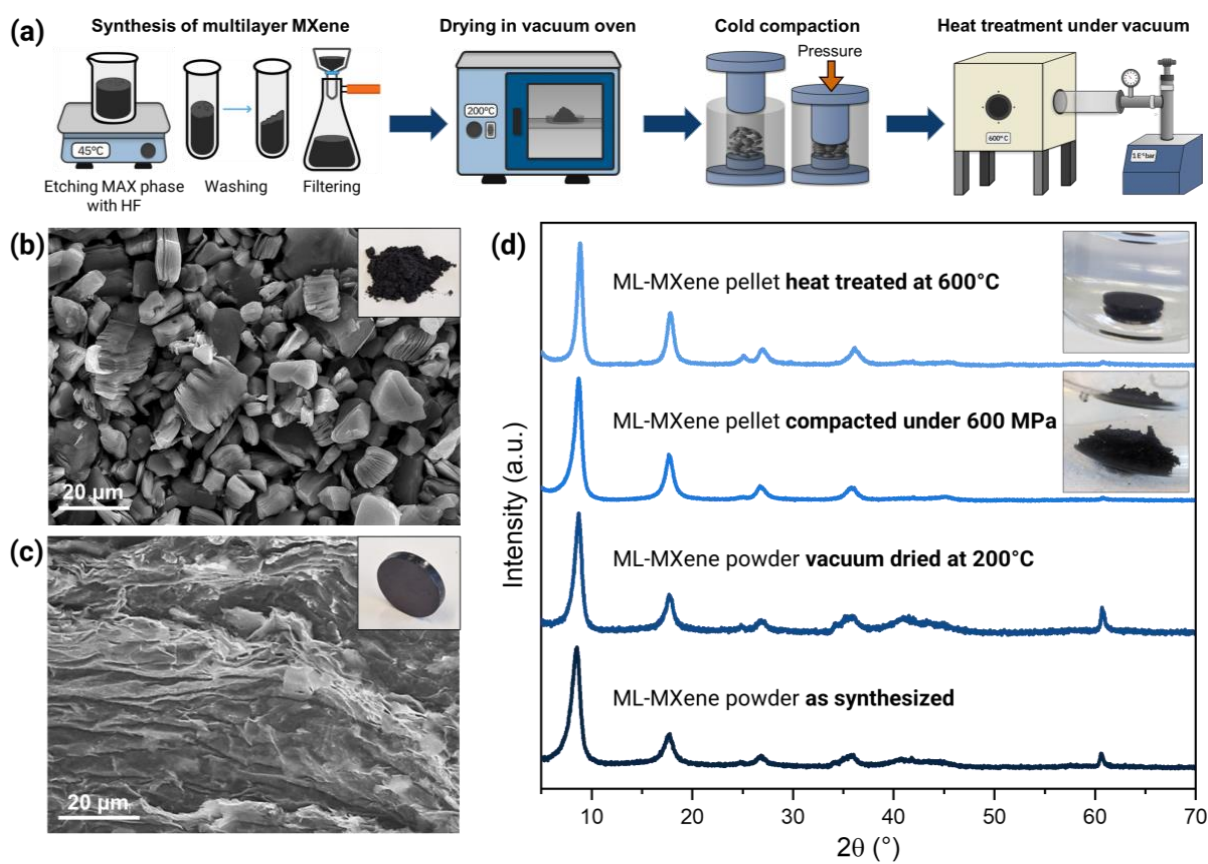

**Figure 1. Preparation of stable, self-supported, bulk 3D MXene electrodes.** (**a**) Schematic illustration of multilayer MXene electrodes preparation process; from synthesis to compaction and heat treatment (HT). SEM images of (**b**) multilayer MXene powder (**c**) cross section of compressed and HT MXene pellet, with representative photographs inserted respectively. (**d**) XRD patterns of the multilayer MXene at various preparation steps, namely directly after synthesis, after initial drying in vacuum

furnace, after compaction into bulk pellets, and after HT. Inserted photographs show the cold compacted pellet disintegrated in 1 M KOH (after ~30 s in solution) in comparison to the stable HT pellet in the same solution and conditions.

Thermal desorption spectroscopy (TDS) spectra confirmed that only water desorbed from compacted MXene pellet during the heating process, showcasing three desorption peaks (**Fig. 2a**). The first peak results from the weakly bound water, evaporating at relatively low temperatures (<150 °C)[29,31], while the second peak or shoulder at higher temperatures (<250 °C) stems from water bound to −OH or other termination groups[32]. Lastly, the third peak (500–600 °C) can be attributed to confined interlayer water[33]. In-situ XRD performed during heating up to 600 °C (**Fig. 2b**) shows the (002) peak shift towards higher 2θ with temperature, indicating a shrinkage in the MXene interlayer d-spacing (**Fig. S4.**). This can be directly correlated to the removal of interlayer water[33], similar to the trend seen in the ex-situ results (**Fig. 1d**). This is further corroborated by in-situ X-ray photoelectron spectroscopy (XPS), showing the corresponding elimination of signal from binding energies associated with $H_2O$ in the O1s spectra (**Fig. 2c**). Additionally, surface terminations remain unchanged after HT (**Fig. S5.**). Note that the spectra observed at 600 °C in the XRD and XPS remain the same after cooling to room temperature.

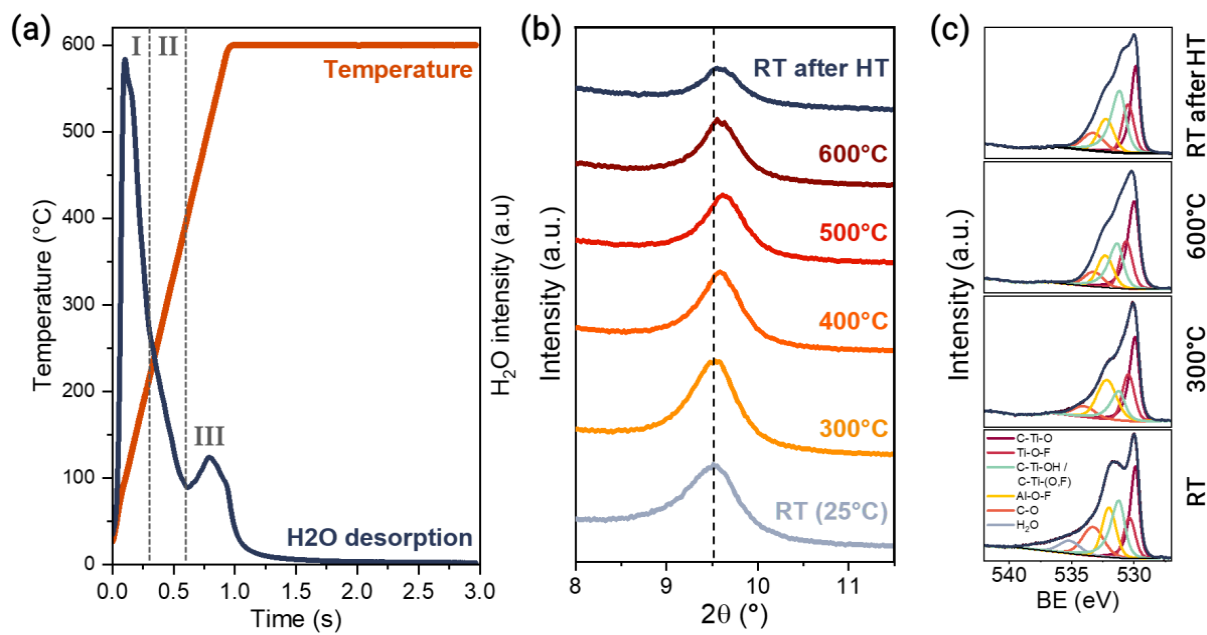

**Figure 2. In-situ characterization of compacted MXene pellets simulating the HT conditions (up to 600 °C).** (**a**) Thermal desorption spectra depicting the thermally driven water removal profile. Three identified peaks are denoted I, II, and III. (**b**) XRD patterns of the MXene basal plane (002) reflection, showcasing the decrease in d-spacing due to removal of interlayer water. (**c**) In-situ high-temperature XPS spectra of O1s binding energies showing the elimination of the water signal.

The electrochemical performance of the resulting pellets was tested using a standard three-electrode cell. Hydrogen evolution reaction (HER) activity was evaluated under acidic conditions, in 1 M $H_2SO_4$ by linear sweep voltammetry (LSV). The overpotential at current densities of -10 mA·cm$^{-2}$ ($\eta_{10}$) and 100 mA cm$^{-2}$ ($\eta_{100}$) were assessed to be 0.2 V and 0.6 V respectively. Although activity behavior is far from common noble metal catalysts, such as Pt, the self-supported $Ti_3C_2T_z$ showcases higher activity (lower overpotential) than reported for other ML-MXene counterparts[34,35], while also achieving higher current densities, as seen in **Fig. 3a**. It is known that pure ML-MXene does not exhibit adequate catalytic capabilities and herein is only used as a proof-of-concept demonstration. The activity of the electrodes can be improved upon by using different MXene compositions[10,34] or incorporating them with nanoparticles or other active sites[20,36]. The oxidation potential of the MXene electrodes was estimated by dynamic polarization measurement (**Fig. S6.**) to be 0.44 V, in agreement with previous studies[37]. It should be emphasized that the stabilized MXene vdW solids retain structural integrity in both acidic and alkaline solutions (1 M $H_2SO_4$ and KOH) during extensive LSV measurements (**Fig. 3a** and **Fig. S6.**).

Long-term stability of the MXene electrodes under HER conditions was evaluated by chronoamperometry (CA) and chronopotentiometry (CP). Acidic conditions are generally accepted as benchmark environments for the evaluation of HER, thus the long-term performance of the electrodes was assessed in 1 M $H_2SO_4$. CA measurements (**Fig. 3b**) were performed at a potential which correlates approximately with a current density of -10 mA·cm$^{-2}$ for 12 h. The current density remained stable, showing only a minor drift (<5%). This small variation indicates that the electrodes maintain their electroactive surface and HER activity without significant deactivation. Complementary CP measurements (**Fig. 3c**) demonstrate stable operation of the electrodes, after an initial activation, with minor potential increase, the insert and **Video S1** shows relatively small bubble formation during CP, promoting HER while preventing excessive bubbles which may lead to contact loss between the surface and electrolyte and unstable performance. The stability observed during CA and CP also suggests the MXene electrodes could withstand processes of long-term electrochemical hydrogen charging, serving as hydrogen storage materials. XPS performed directly after the stability tests shows only minor superficial oxidation (**Fig. S7.**).

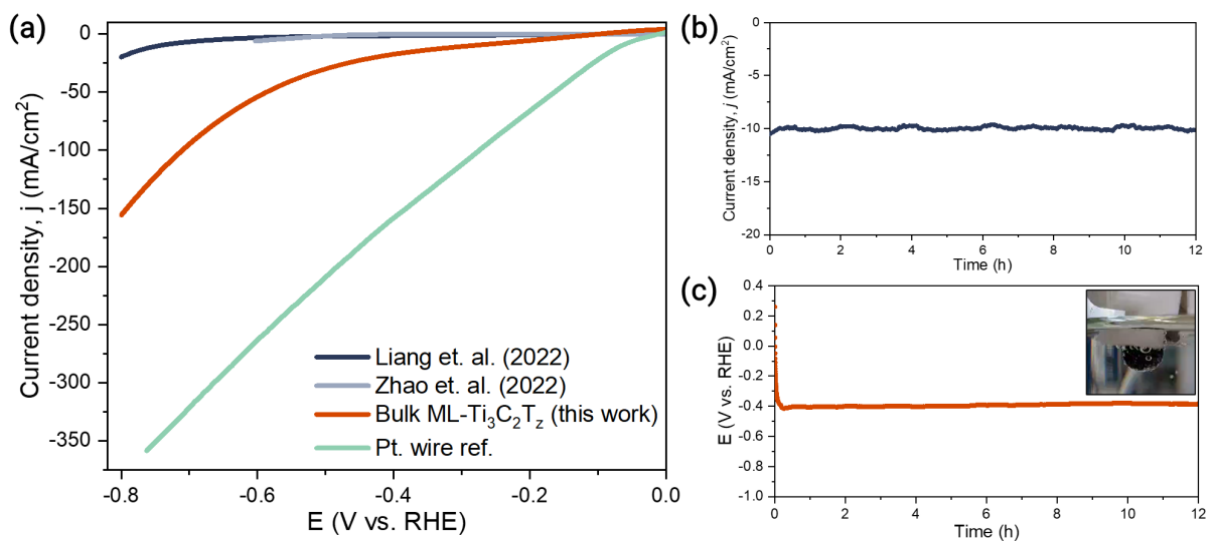

**Figure 3. Electrochemical performance of bulk MXene electrodes under HER conditions.** (**a**) LSV curve for HER measured in 1 M $H_2SO_4$ aqueous solution for Bulk ML-$Ti_3C_2T_z$ electrodes and Pt wire, in comparison to state-of-the-art ML-MXene[34,35]. Stability tests for HER presenting minimal changes in electrode stability (**b**) Chronoamperometry (CA) curve for HER under a potential of 0.4 V (**c**) Chronopotentiometric (CP) curve with the constant current density of 10 mA·cm$^{-2}$, insert shows MXene electrode under CP test conditions and small bubble formation during testing.

In summary, we developed an effective method for the production of self-supported, bulk vdW MXene electrodes with mechanical robustness and electrochemical stability suitable for hydrogen related applications. By compacting ML-MXene under high pressure (~600 MPa) and applying a vacuum heat treatment at 600 °C to remove confined interlayer water, we obtained bulk 3D MXene electrodes that remain stable in acidic, neutral, and alkaline environments and under electrochemical operation. These MXene electrodes demonstrate consistent HER activity, while maintaining their performance during long-term CA and CP cycling, with negligible oxidation or change in termination chemistry. This work establishes the first demonstration of chemically and mechanically stable bulk MXene electrodes and provides a generalizable pathway for integrating MXenes or MXene-based heterostructures into practical hydrogen evolution and hydrogen storage systems. This approach is simple, cost-effective, and scalable, offering a promising route for advancing MXenes toward real-world electrochemical energy technologies.

**Experimental section**

*Synthesis of $Ti_3AlC_2$ MAX Phase:* TiC (Alfa Aesar, 99.5% 2 µm), Al (Alfa Aesar, 99.5%, 325 mesh), and Ti (Alfa Aesar, 99.5%, 325 mesh) were mixed in a molar ratio of 2:1.15:1 respectively. The powders were mixed in a tumbler ball mixer at 100 rpm for 24 h and heat-treated at 1400 °C for 3 h under a flowing Ar/5%$H_2$. The heating and cooling rates were 5 °C/min. The resulting block was ground to a fine powder and filtered through a 400 mesh (<38 µm) sieve, yielding a powder consisting of particles with a size around 20 µm.

*Synthesis of $Ti_3C_2T_z$ MXene:* 1 g of $Ti_3AlC_2$ powder (particle size < 20 µm) was etched for 48 h at 45 °C in 20 mL of a solution with a 6:3:1 ratio of HCl:$H_2O$:HF. The resulting sediment was washed with 1 M HCl, followed by repeated centrifugation at 3500 rpm for 2 min with DI until reaching a neutral pH. The resulting multilayer MXene powder was vacuum filtered, then dried for 4 h at 80 °C, followed by 20 h at 200 °C under vacuum.

*MXene pellet preparation:* $Ti_3C_2T_z$ MXene powder was compacted by using a hydraulic press and steel die with diameters of 5-10 mm. Compaction was performed under roughly 300, 400 and 600 MPa. The resulting pellets were heat-treated at 600 °C (10 °C/min heating rate) for 2 h under vacuum, and for comparison also at the same heating conditions under 100% Ar protective atmosphere.

*Material Characterizations: X-ray diffraction (**XRD**)* measurements of $Ti_3C_2T_z$ powder and pellets were performed on a Malvern Panalytical Aeris diffractometer with a Bragg–Brentano geometry and a Cu Kα (λ = 1.5406 Å) radiation source. Diffraction patterns were recorded at 2θ = 3-85°, at increments of 0.02°. *Scanning electron microscopy (**SEM**)* characterization was performed using a ZEISS Sigma microscope, the acceleration voltage and beam current were 15 kV and 9.5 nA, respectively. ***In-situ XRD*** was performed using a Rigaku Smartlab 9kW diffractometer with a Bragg–Brentano geometry and a Cu Kα (λ = 1.5406 Å) radiation source, probing 2θ = 3 to 30°, at increments of 0.02°. The $Ti_3C_2T_z$ specimen (10 mm diameter) was placed on a Macor sample stage in an Anton Paar XRK-900 reaction chamber. The sample was in 100% $N_2$ protective atmosphere, and patterns were collected for temperatures ranging from RT to 600 °C (heating rate of 10 °C/min). *X-ray photoelectron spectroscopy (**XPS**)* measurements were performed using an AXIS SUPRA spectrometer (Kratos Analytical Ltd., UK) with a monochromatic Al Kα source and a hemispherical analyzer. For high-resolution scans, a pass energy of 20 eV, a step size of 0.05 eV and dwell time of 200 ms were used and 10 to 15 sweeps were conducted. Wide scans, used for quantification, were obtained with a pass energy of 160 eV, step size of 0.250 eV, dwell time of 100 ms and 5 sweeps. In-situ heating was performed at 300 °C and 600 °C, where the temperature was reached employing a heating rate of 10 K/min. The temperature was held constant for the duration of each measurement, lasting approximately 2.5 h. After cooling overnight, an additional measurement of the cooled sample was performed. The spectra were fitted with the CasaXPS software with Tougaard background subtraction. For the MXene related peaks asymmetric lineshapes are assumed while all other peaks are approximated with symmetric Gaussian/Lorentzian line shapes. *Thermal desorption spectroscopy* (**TDS**) measurements were done on a Hiden TPD Workstation. The cell went into vacuum in ~20 min, starting the measurement with a heating rate of 10 °C/min from room temperature to 600 °C, and a holding time of 2 h – accurately replicating the heat treatment conditions.

*Electrochemical characterizations:* All electrochemical measurements were performed using a Biologic VSP-300 potentiostat, in a 3-electrode cell configuration using RHE reference electrode and Pt mesh counter electrode, in a solution of 1 M $H_2SO_4$. The $Ti_3C_2T_z$ specimen served as the working electrode and was held by a PTFE Pt electrode holder. Linear sweep voltammetry (LSV) was carried out in the potential window from 0 to –0.8 V vs. RHE at a scan rate of 0.01 mV·s$^{-1}$. Chronoamperometry stability test was performed at a potential of 0.4 V, correlating with *j* of approximately 10 mA·cm$^{-2}$. All tests were carried out in ambient conditions.

## Acknowledgements

This work has been performed in the framework of the International Max Planck Research School for Sustainable Metallurgy (IMPRS SusMet). We thank Dr. Lee Shelly from the Ilse Katz Institute for Nanoscale Science & Technology at Ben-Gurion University of the Negev for performing the BET measurements; Benjamin Breitbach for support with XRD measurements; Zakarya Meshou for overall laboratory support and Tommy Miyar for schematic illustrations.